\documentclass[vecphys,fleqn]{svmult}

\usepackage{makeidx}         
\usepackage{graphicx}
\usepackage{multicol}
\usepackage[bottom]{footmisc}

\begin{document}

\title*{Quantitative Treatment of Decoherence}

\author{{\bf Leonid Fedichkin}\ and\ {\bf Vladimir Privman}}
\institute{Center for Quantum Device Technology,
Department of Physics and{}\\{}Department of Electrical and Computer Engineering,
Clarkson University, Potsdam, New York 13699--5721, USA
}
\maketitle


\section*{Abstract}

We review several approaches to define and quantify decoherence. We find that a measure based on a norm of deviation of the density matrix is appropriate for quantifying decoherence for quantum registers. For a semiconductor double quantum dot charge qubit, evaluation of this measure is presented. For a general class of decoherence processes, including those occurring in semiconductor qubits, we establish that this measure is additive: It scales linearly with the number of qubits in the quantum register.

\section{Introduction}

Decoherence~\cite{open,CL,Chakravarty,Grabert,vanKampen,nonMarkov,Anastopoulos,Ford,Braun,Lewis,Wang,Lutz,Khaetskii,OConnell,Strunz,Haake,PMV,short,Privman}
is an important physical phenomenon
occurring inevitably
in most experiments dealing with quantum objects.
It is usually defined as
a process whereby the
physical system of interest interacts
with environment or other larger system
with complex structure
and, because of this interaction, changes
its evolution from unperturbed, coherent internal dynamics.
In some sense, the information about the initial
and subsequent states of system undergoing decoherence is
leaking into the outer world: The system is no longer described by a wave function,
but rather by the statistical density
matrix~\cite{Neumann,Abragam,therm2,therm1,Louisell}.
The quantum wave
function description
only applies to the total system, including the environmental modes,
which has much more degrees of freedom.
Because of the importance of quantum coherence for quantum information processing~\cite{Shnirman,Ekert,Kventsel,Kane,Loss,Imamoglu,Rossi,Nakamura,Tanamoto,Platzman,%
Sanders,Burkard,Vrijen,Fedichkin,Bandyopadhyay,Larionov,%
Hawrylak,Alex1,Alex2,Openov,Hayashi,Fujisawa,%
Dzurak1,Dzurak2,Cain,Smith,Barrett,Ahn,Ben1,Ben2,%
qec,Steane,Bennett,Calderbank,SteanePRA,Gottesman,Knill,Kitaev,%
Kitaev2,Kitaev3,Preskill,DiVincenzo},
quantitative characterization of decoherence has become an active research
field with many open problems.

Since quantum information processing
requires maintaining high level of coherence,
emphasis has recently shifted from large-time
system dynamics at experimentally
better studied coherence-decay time scales
to almost perfectly coherent dynamics at much shorter times.
Many quantum systems proposed as candidates
for qubits (quantum bits) for practical realizations of
quantum computing require quantitative evaluation of their coherence.
In other words, a single measure characterizing
decoherence is desirable for comparison of different
qubit designs and their optimization.
Besides the evaluation of single qubit performance
one also has to analyze scaling of decoherence
as the register size (the number of qubits involved)
increases. Direct quantitative calculations
of decoherence of even few-qubit quantum registers
are not feasible.
Therefore, a practical approach has been to explore
quantitative single-parameter measures of decoherence~\cite{norm},
develop techniques to calculate such measures at least
approximately for realistic one- and two-qubit systems~\cite{dd,dd_ieee},
and then establish scaling (additivity~\cite{additivity,jctn}) for multi-qubit quantum systems.

In Section \ref{Secti2}, we outline different approaches to define and quantify
decoherence. We argue that a measure based on a properly defined norm of deviation of the density matrix
is appropriate for quantifying decoherence in quantum registers.
For a semiconductor double quantum dot qubit, evaluation of this measure is reviewed in Section \ref{Secti3}.
For a general class
of decoherence processes, including those occurring in semiconductor qubits considered in Section \ref{Secti3},
we argue, in Section \ref{Secti4}, that
this measure is additive. Thus, the level of quantum noise scales linearly with the number of qubits.

\section{Measures of Decoherence}\label{Secti2}

In this section, we consider briefly several approaches to quantifying the degree
of decoherence due to interactions with environment. In Subsection \ref{Subsecti21}, we
discuss the approach based on the asymptotic relaxation time scales. The entropy and
idempotency-defect measures are reviewed in Subsection \ref{Subsecti22}. The fidelity measure of
decoherence is considered in Subsection \ref{Subsecti23}. In Subsection \ref{Subsecti24}, we
review our results on the operator norm measures of decoherence.
Subsection \ref{Subsecti25} discusses an approach to eliminate the initial-state
dependence of the decoherence measures.

\subsection{Relaxation Time Scales}\label{Subsecti21}
Decoherence of quantum systems is frequently
characterized by the asymptotic rates at which
they reach thermal equilibrium at temperature $T$. One of the reasons for focusing on
relaxation rates is that large-time behavior
is relatively easy to observe in ensemble experiments.
Markovian approximation schemes typically yield exponential
approach to the limiting values of the density matrix elements for
large times \cite{Abragam,therm2,therm1}. For a two-state
system, this defines the time scales $T_1$ and
$T_2$, associated, respectively, with the approach by the diagonal
(thermalization) and off-diagonal (dephasing, decoherence) density-matrix
elements to their limiting values. More generally, for large times we
approximate deviations from stationary values of diagonal and
off-diagonal density matrix elements as
\begin{equation}
 \rho _{kk} (t) - \rho _{kk}(\infty) \propto e^{ - t/T_{kk} } ,
\end{equation}\par\noindent
\begin{equation}
 \rho _{jk} (t) \propto e^{ - t/T_{jk} }  \qquad (j \ne k) .
\end{equation}\par\noindent
The shortest time among $T_{kk}$ is
often identified as $T_1$. Similarly, $T_2$ can be defined as the
shortest time among $T_{n \ne m}$. These definitions yield the
characteristic times of thermalization and decoherence (dephasing).

For systems candidate for quantum computing realizations,
noise effects are commonly reduced by working at very low temperatures
and making their structure features nanosize for strong quantization.
Then for the decoherence and thermalization times we have,
$T_2  \ll T_1 $, e.g., \cite{Abragam}. Therefore, the
decoherence time is a more crucial parameter for quantum computing
considerations. The time scale $T_2$ is compared to the ``clock''
times of quantum control, i.e., the quantum gate functions,
$T_g$, in order to ensure the fault-tolerant error correction criterion
$T_g/T_2 \leq O\left( 10^{-4}\right)$, e.g., \cite{DiVincenzo}.

The disadvantages of this type of
analysis are that the exponential behavior of the density matrix elements
in the energy basis is applicable only for large times, whereas
for quantum computing applications, the short-time behavior
is usually relevant \cite{short}.
Moreover, while the energy basis is natural
for large times, the choice of the preferred basis is not obvious for
short and intermediate times \cite{short,basis}. Therefore, the
time scales $T_1$ and $T_2$ have limited applicability in evaluating quantum
computing scalability.

\subsection{Quantum Entropy}\label{Subsecti22}
An alternative approach is to calculate the entropy \cite{Neumann}
of the system,
\begin{equation}
S(t)=- {\rm Tr}\left(  \rho \ln  \rho \right),
\end{equation}\par\noindent
or the idempotency defect, also termed the first order entropy \cite{Kim,Zurek,Zagur},
\begin{equation}
 \label{trace}
s(t)=1 - {\rm Tr} \left( \rho ^2 \right).
\end{equation}\par\noindent
Both expressions are basis independent, have a minimum at pure
states and effectively describe the degree of the state's
``purity.'' Any deviation from a pure state leads to the
deviation from the minimal values, 0, for both measures,
\begin{equation}
S_{\,\rm pure\ state}(t)= s_{\,\rm pure\ state}(t)= 0.
\end{equation}

\subsection{Fidelity}\label{Subsecti23}

Writing the total Hamiltonian as follows,
\begin{equation}\label{f0}
  H=H_S+H_B+H_I,
\end{equation}\par\noindent
where $H_S$ is the term describing internal system dynamics,
$H_B$ governs the evolution of environment,
and $H_I$ describes system-environment interaction,
let us now define the fidelity \cite{Dalton,Fidelity2},
\begin{equation}\label{f1}
F(t)={\rm Tr}_{\,S}  \left[ \, \rho _{\rm ideal}(t) \, \rho (t) \, \right].
\end{equation}\par\noindent
Here the trace is over the system degrees of freedom, and $\rho_{\rm ideal}(t)$
represents the pure-state evolution of the system under $H_S$ only,
without interaction with the environment ($H_I=0$).
In general, the Hamiltonian term $H_S$ governing the system dynamics can be time dependent.
For the sake of simplicity throughout this review we restrict our analysis by
constant $H_S$ since approximate evaluation of decoherence
can be done for qubits controlled by constant Hamiltonian.
In this case
\begin{equation}\label{f2}
\rho _{\rm ideal}(t)= e^{-iH_S t}\rho(0)\, e^{iH_S t}.
\end{equation}
More sophisticated  scenarios
with qubits evolving under time dependent $H_S$ were considered in~\cite{Brandes,Brandes2,Solenov}.

The fidelity provides a certain measure of decoherence in terms of
the difference between the ``real,'' environmentally influenced,
$\rho (t)$, evolution and the ``free'' evolution, $\rho_{\rm
ideal} (t)$. It will attain its maximal value, 1, only provided
$\rho (t) = \rho_{\rm ideal} (t)$. This property relies on the
fact the $ \rho_{\rm ideal} (t)$ remains a projection operator
(pure state) for all times $t \geq 0$.

As an illustrative example consider a two-level system decaying from the excited to
ground state, when there is no internal system dynamics,
\begin{equation}
\rho _{\rm ideal} (t) =\left(
\begin{array}{cc}
0 & 0 \\
0 & 1
\end{array}
\right),
\end{equation}\par\noindent
\begin{equation}
\rho (t)=\left(
\begin{array}{cc}
1-e^{-\Gamma t} & 0 \\
0 & e^{-\Gamma t}
\end{array}
\right),
\end{equation}\par\noindent
and the fidelity is a monotonic function of time,
\begin{equation}\label{f3}
F(t)=e^{-\Gamma t}.
\end{equation}\par\noindent

Note that the requirement that $\rho_{\rm ideal}(t)$ is a pure-state (projection operator),
excludes, in particular, any $T>0$ thermalized state as the initial system state.
For example, let us consider the application of the fidelity measure
for the infinite-temperature initial state of our two level system. We have
\begin{equation} \rho (0)=\rho _{\rm ideal}(t)=\left(
\begin{array}{cc}
1/2 & 0 \\
0 & 1/2
\end{array}
\right),
\end{equation}\par\noindent
which is not a projection operator. The spontaneous-decay density matrix is then
\begin{equation}
\rho (t)=\left(
\begin{array}{cc}
1-(e^{-\Gamma t}/2) & 0 \\
0 & e^{-\Gamma t}/2
\end{array}\right).
\end{equation}\par\noindent
The fidelity remains constant
\begin{equation}\label{f3.1}
F(t)=1/2,
\end{equation}\par\noindent
and it does not provide any information of the time dependence of the decay process.

\subsection{Norm of Deviation}\label{Subsecti24}

In this subsection we consider the operator norms \cite{Kato}
that measure the deviation of the system from the ideal state, to
quantify the degree of decoherence as proposed in~\cite{norm}.
Such measures do not require the
initial density matrix to be pure-state. We define the deviation
according to
\begin{equation}\label{deviation}
  \sigma(t)  \equiv   \rho(t)  -   \rho_{\rm ideal} (t)  .
\end{equation}\par\noindent
We can use, for instance, the
eigenvalue norm,
\begin{equation}\label{n11}
\left\|\sigma \right\|_{\lambda} = {\max_i} \left|
{\lambda _i } \right|,
\end{equation}\par\noindent
or the trace norm,
\begin{equation}\label{tracenorm}
\left\|   \sigma  \right\|_{{\rm Tr}}  = \sum\limits_i {\left|
{\lambda _i } \right|},
\end{equation}\par\noindent
etc., where $\lambda_i$ are the eigenvalues of the deviation
operator (\ref{deviation}).
A more precise definition of the eigenvalue norm for a
linear operator, $A$, is \cite{Kato}
\begin{equation}\label{n2}
\left\| A \right\| = \mathop {\sup }\limits_{\varphi  \ne 0}
\left[ {\frac{{ \langle \varphi |A^ \dagger  A|\varphi  \rangle }}{{
\langle \varphi |\varphi  \rangle }}} \right]^{1/2}.
\end{equation}\par\noindent
Since density operators are bounded, their norms, as well
the norm of the deviation, can be always evaluated. Furthermore,
since the density operators are Hermitian, this definition
obviously reduces to the eigenvalue norm (\ref{n11}). We also note that $\left\| A \right\|=0$ implies that $A=0$.

The calculation of these norms is sometimes simplified by the observation that
$\sigma(t)$ is traceless. Specifically, for two-level systems, we get
\begin{equation} \left\|   \sigma
\right\|_{\lambda} = \sqrt {\left| {\sigma _{00} } \right|^2
+ \left| {\sigma_{01} } \right|^2 } = {1 \over 2}
\left\|   \sigma  \right\|_{{\rm Tr}}.
\end{equation}\par\noindent
For our example of the two-level system undergoing spontaneous decay,
the norm is
\begin{equation}
\left\|   \sigma
\right\|_{\lambda} = 1 - e^{-\Gamma t} .
\end{equation}\par\noindent

\subsection{Arbitrary Initial States}\label{Subsecti25}

The measures considered in the preceding subsections quantify decoherence of a system
provided its initial state is given. However, this is not always the
case. In quantum computing, it is impractical to keep track of all the possible initial states for each quantum register, that might be needed for implementing a particular quantum algorithm. Furthermore, even the preparation of the initial state can introduce additional noise. Therefore, for evaluation of fault-tolerance (scalability), it will be necessary to obtain an upper-bound estimate of decoherence for an arbitrary initial state.

To characterize decoherence for an arbitrary initial state, pure or mixed,
we proposed~\cite{norm} to use the maximal norm, $D$, which
is determined as an operator norm maximized over all initial density matrices.
It is defined as the worst case scenario error,
\begin{equation}\label{normD}
  D(t) = \sup_{\rho (0)}\bigg(\left\| \sigma (t,\rho (0))\right\|_{\lambda}  \bigg).
\end{equation}\par\noindent

For realistic two-level systems coupled to
various types of environmental modes,
the expressions of the maximal norm are surprisingly
elegant and compact. They are
usually monotonic and contain no oscillations due to the
internal system dynamics, as, for example, are the results
obtained for semiconductor quantum dot qubits considered in the next section.

In summary, we have considered several approaches to
quantifying decoherence: relaxation times, entropy and fidelity measures,
and norms of deviation, and we defined the maximal measure that is not
dependent on the initial state, and which will be later shown to be additive; see Section~\ref{Secti4}.

\section{Decoherence of Double Quantum Dot Charge Qubits}\label{Secti3}

As a representative example, let us review evaluation of decoherence for semiconductor quantum dots.
Quantum devices based on solid-state
nanostructures have been among the major candidates for
large-scale quantum computation because they can draw on existing
advances in nanotechnology and materials
processing~\cite{roadmap}. Several designs of semiconductor quantum
bits (qubits) were proposed~\cite{Kventsel,Kane,Vrijen,Hawrylak,Alex1,Alex2,Openov,solid1,solid2,Barenco}.
In particular, the encoding of quantum information into spatial degrees of
freedom of electron placed in a quantum dot was considered
in~\cite{Hawrylak,Alex1,Alex2,Openov,Barenco}. A relatively fast
decay of coherence of electron states in ordinary quantum dots,
e.g., \cite{Ekert}, can be partially suppressed by
encoding quantum information in a subspace of electron states in
specially designed arrays of quantum dots (artificial crystals),
proposed in \cite{Zanardi}. Actually, under
certain conditions even double-dot systems in semiconductors can
be relatively well protected against decoherence due to
their interactions with phonons and electromagnetic
fields~\cite{Fedichkin}. This observation was confirmed in recent
experiments \cite{Hayashi}, which demonstrated
coherent quantum oscillations of an electron in a
double-dot structure.

Several designs of double-dot qubits have been explored in recent
experiments \cite{Fujisawa,Dzurak1,Dzurak2,Cain,Smith}
carried out at temperatures ranging from tens and hundreds of mK.
Temperature dependence of relaxation rates in Si charge qubits was studied
theoretically in \cite{Barrett,Ahn}. Recently, it has been pointed out \cite{dd} that
in the zero-temperature limit and for
conventional double-dot structures higher order
processes in electron-phonon interaction dominate decoherence.

In this work, we studied the acoustic phonon bath
as the main source of decoherence for the considered type of qubit,
which is supported by theoretical and experimental evidence,
e.g., \cite{Fedichkin,Fujisawa}. Decoherence due to different sources, e.g.,
due to trapping center defects \cite{Pashkin,Martin},
can play important role in other situations.

In the next subsection, we outline the structure of double-dot qubits.
Subsections \ref{Piezo} and \ref{Deform} are devoted to the consideration of the
electron-phonon interaction for two
realistic cases: In Subsection \ref{Piezo} we analyze the piezoacoustic
interaction in crystals with zinc-blende lattice and with parabolic
quantum dot confinement potential. Double-dots with prevalence of
piezointeraction have been fabricated \cite{Hayashi} in gated GaAs/AlGaAs
heterostructures. In Subsection \ref{Deform} we study the
deformation interaction with acoustic phonons in ``quantum dots''
formed by double-impurities in semiconductors with inversion
symmetry of elementary lattice cell. Experiments with the latter
type of double-dot systems have been reported in \cite{Dzurak1,Dzurak2}.
Finally, Subsections \ref{Subsecti34}, \ref{NOT}, \ref{PI} and \ref{Subsecti37} present
illustrative calculations of the noise level for selected quantum gates.

\subsection{Model}\label{Susecti31}

\begin{figure}[t]
\includegraphics[width=11 cm] 
{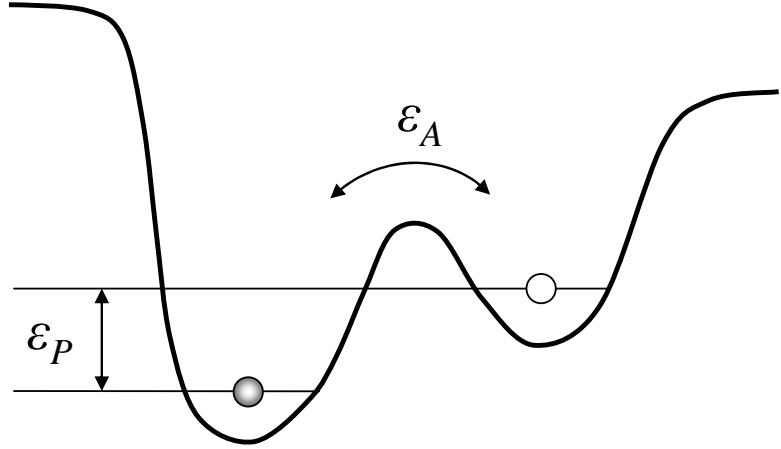} \caption{Electron in a double well potential.}
\label{fig:1}
\end{figure}

We consider a double-dot structure sketched in Figure \ref{fig:1}.
It consists of two quantum
dots coupled to each other via a tunneling barrier and containing a single
electron hopping between the dots. We limit our
consideration to double-dot structures in which the energy required to
transfer to the upper levels is much higher than the lattice temperature
and energy spacing between the two lowest levels.

The electron is considered to be in a
superposition of two basis states,  $|0\rangle$ and $|1\rangle$,
\begin{equation}
\psi  = \alpha \psi_0 + \beta \psi_1 .
\end{equation}
The states that define the ``logical'' basis are not the physical
ground and first excited state of the double-dot system.
Instead, $\psi_0$ (the ``0'' state of the qubit) is chosen to be
localized at the first quantum dot and, to a zeroth order approximation,
be similar to the ground state of that dot if it were isolated.
Similarly,  $\psi_1$ (the ``1'' state) resembles the ground state
of the second dot (if it were isolated).
This assumes that the dots are sufficiently (but not necessarily exactly) symmetric.
We denote the
coordinates of the potential minima  of the dots (dot centers) as
vectors $\mathbf R_0$ and $\mathbf R_1$, respectively. The distance
between the dot centers is
\begin{equation}
L\equiv |\mathbf L|\equiv|\mathbf
R_1-\mathbf R_0| .
\end{equation}

The Hamiltonian of an electron within a phonon
environment is given by
\begin{equation} \label{H}
 H = H_e + H_p + H_{ep}.
\end{equation}
The electron term is
\begin{equation}\label{He}
 H_e = - \frac{1}{2}\varepsilon _{A}(t)\sigma _x- \frac{1}{2}\varepsilon _{P}(t)\sigma _z,
\end{equation}
where $ \sigma_x$ and $\sigma _z$ are Pauli matrices, whereas
 $\varepsilon _{A}(t)$ and $\varepsilon _{P}(t)$ can have time-dependent, as determined by unitary single-qubit quantum gate-functions that are carried out. They can
be controlled externally by adjusting the potential on the
control electrodes (gates) surrounding the double-dot system. For constant $\varepsilon _{A}$ and $\varepsilon _{P}$, the energy splitting
between the electron energy levels is
\begin{equation}
 \varepsilon=\sqrt{\varepsilon _{A}^2+\varepsilon _{P}^2}.
\end{equation}

The Hamiltonian of the phonon bath is described by
\begin{equation}
 H_p = \sum\limits_{\mathbf{q},\lambda } \hbar \omega_q {\kern 1pt} b_{\mathbf{%
q},\lambda }^{\dagger}  b_{\mathbf{q},\lambda },
\end{equation}
where $ b^{\dagger}_{\mathbf{q},\lambda}$ and $
b_{\mathbf{q},\lambda}$ are, respectively, the creation and
annihilation operators of phonons characterized by the
wave vector $\mathbf q\,$ and polarization $\lambda$. We
approximately assume isotropic acoustic
phonons, with a linear dispersion,
\begin{equation}\omega_q = s q,\end{equation}
where $s$
is the speed of sound in the semiconductor material. In the next subsection we
show that the electron-phonon interaction can be derived in the form
\begin{equation}  \label{int}
 H_{ep}=\sum\limits_{\mathbf{q},\lambda }
 \sigma_z {\kern 1pt}\left( g_{\mathbf{q},\lambda }  b_{\mathbf{q},
\lambda}^{\dagger} +  g_{\mathbf{q},\lambda }^* b_{\mathbf{q},
\lambda}\right),
\end{equation}
with the coupling constants $g_{\mathbf{{q}, \lambda}}$ determined by the architecture of the double-dot and the properties of the material crystal structure.

\subsection{Piezoelectric Interaction}\label{Piezo}

The derivation in this subsection follows \cite{dd,dd_ieee}. The piezoacoustic electron-phonon
interaction \cite{Mahan} is described by
\begin{equation}  \label{p1}
H_{ep}=i{\sum\limits_{\mathbf q,\lambda} }\left( \frac \hbar
{2\rho s q V }\right) ^{1/2}\!\! M_\lambda ({\mathbf q})F (\mathbf
q)(b_{\mathbf q}+b_{-\mathbf q}^{\dagger}).
\end{equation}
Here $\rho$ is the density of the semiconductor material, $V$ is the volume of semiconductor,
 and for the matrix element
$M_\lambda ({\mathbf q})$, one can derive
\begin{equation}  \label{p2}
M_\lambda ({\mathbf q})=\frac 1{2q^2}{\sum\limits_{ijk} }(\xi
_i^{\vphantom{Q}}q_j^{\vphantom{Q}}+\xi _j^{\vphantom{Q}}q_i^{\vphantom{Q}})q_k^{\vphantom{Q}} M_{ijk}.
\end{equation}
In this expression, $\xi _j$ are the polarization vector components for polarization $\lambda$, while $M_{ijk}$ express the electric field as a linear response to the stress,
\begin{equation}  \label{p3}
E_k={\sum\limits_{ij} }M_{ijk}S_{ij}.
\end{equation}
For a crystal with zinc-blende lattice, exemplified by GaAs, the tensor $M_{ijk}$ has
only those components non-zero for which all three indexes $i$, $j$, $k$ are
different; furthermore, all these components are equal $M_{ijk}=M$. Thus, we have
\begin{equation}  \label{p4}
M_\lambda (\mathbf q)=\frac M{q^2}(\xi^{\vphantom{Q}}_1q_2^{\vphantom{Q}}q_3^{\vphantom{Q}}+ \xi^{\vphantom{Q}}_2 q_1^{\vphantom{Q}}q_3^{\vphantom{Q}}+\xi^{\vphantom{Q}}_3q_1^{\vphantom{Q}}q_2^{\vphantom{Q}}).
\end{equation}

The form factor $F(\mathbf{{q})}$ accounting for that the electrons in the quantum dot
geometry are not plane waves, is
\begin{equation}\label{e-density}
F (\mathbf q)=
\sum\limits_{j,k}%
c_j^{\dagger}c_k\int d^3r \phi_j^{*}(\mathbf r)\phi_k(\mathbf
r)e^{-i\mathbf q\cdot \mathbf r},
\end{equation}
where $c_k$, $c^{\dagger }_j$ are annihilation and creation
operators of the basis states $k,j=0,1$. For
gate-engineered quantum dots, we consider the ground states in each dot to have an approximately Gaussian shape
\begin{equation}  \label{1}
\phi_j(\mathbf r)=\frac 1{a^{3/2}\pi
^{3/4}}\,e^{-|\mathbf r-\mathbf R_j|^2/2a^2},
\end{equation}
where $2a$ is a characteristic size of the dots.

We assume that the distance between the dots, $L$, is sufficiently
large compared to $a$, to ensure that the different dots wave functions do not
overlap significantly,
\begin{equation}
\bigg| \int d^3r \phi_j^{*}(\mathbf r)\phi_k(\mathbf r)e^{-i \mathbf q\cdot \mathbf r} \bigg| \ll 1, \quad {\rm
for} \quad j\neq k.
\end{equation}
This implies that the coupling leading to tunneling between the dots is small, as is the
case for the recently studied experimental structures \cite{Hayashi,Fujisawa,Dzurak1,Dzurak2}, where the
splitting due to tunneling, measured by
$\varepsilon _{A}$, was
below $20\,\mu$eV, while the electron quantization energy in each dot
was at least several meV.

For $j=k$, we obtain
\begin{eqnarray}\label{3}\nonumber
\int d^3 r \phi_j^{*}(\mathbf r)\phi_j(\mathbf r)
e^{-i\mathbf q\cdot \mathbf r}=
\frac{1}{a^3\pi^{3/2} }\int d^3 r
e^{-|{\mathbf r} -{\mathbf R}_j|^2/a^2}e^{-i\mathbf q\cdot \mathbf r}
\end{eqnarray}
\begin{eqnarray}
=e^{-i{{\mathbf q}\cdot {{\mathbf R}_j}}}e^{-a^2q^2/4}.
\end{eqnarray}
The resulting form factor is
\begin{equation}
F (q)=e^{-a^2q^2/4}e^{-i\mathbf q\cdot \mathbf R}(c_0^{\dagger
}c_0e^{i\mathbf q\cdot \mathbf L/2}+c_1^{\dagger }c_1e^{-i\mathbf
q\cdot \mathbf L/2}),
\end{equation}
where $\mathbf R=\left(\mathbf R_0+\mathbf R_1\right)/2$. Finally, we get
\begin{equation}\label{form}
F (q)=e^{-a^2q^2/4}e^{-i\mathbf q\cdot \mathbf R}\left[\cos
(\mathbf q\cdot \mathbf L /2)I+i\sin (\mathbf q\cdot \mathbf L
/2)\sigma _z\right] \label{4},
\end{equation}
where $I$ is the identity operator. Only the second term in (\ref{form}),
which is not proportional to $I$,
represents an interaction affecting the qubit states.
It leads to a Hamiltonian term of the form (\ref{int}),
with coupling constants
\begin{eqnarray}  \label{p6}\nonumber
g_{\mathbf q, \lambda}&=&-\left( \frac{\hbar   }{2 \rho q s V
}\right) ^{1/2}Me^{-a^2q^2/4 - i\mathbf q\cdot \mathbf R}\\
&&\times(\xi_1^{\vphantom{Q}}e_2^{\vphantom{Q}}e_3^{\vphantom{Q}}+
\xi_2^{\vphantom{Q}}e_1^{\vphantom{Q}}e_3^{\vphantom{Q}}
+\xi_3^{\vphantom{Q}}e_1^{\vphantom{Q}}e_2^{\vphantom{Q}})
\sin (\mathbf q\cdot \mathbf L/2),
\end{eqnarray}
 where $e_k=q_k/q$.

\subsection{Deformation Interaction} \label{Deform}

Deformation coupling with acoustic phonons \cite{Mahan} is
described by
\begin{equation}
 H_{ep}=\Xi \sum\limits_{\mathbf q, \lambda} \left( {\frac{\hbar
}{{2\rho qs V}}} \right)^{1/2} \!\! q F( \mathbf q){\kern 1pt}(
b_{\mathbf q,\lambda }^{\dagger} +  b_{-\mathbf q,\lambda }),
\end{equation}
where $\Xi$ is a material-dependent constant termed the ``deformation potential.''

Here we consider a particular double-dot-like
nanostructure which has been a focus of recent experiments, due to advances in its
fabrication \cite{Dzurak1,Dzurak2} by controlled single-ion implantation:
A double-impurity Si structure with Hydrogen-like electron confinement
potentials for at both impurities (P atoms). We consider a
Hydrogen-like impurity state,
\begin{equation}\label{hy1}
\phi _i(r)=\frac 1{a^{3/2}\pi ^{1/2}}\,e^{-|\mathbf r-\mathbf
R_i|/a},
\end{equation}
where $a$ is the effective Bohr radius.  The form factor in this
case is given by the following formula,
\begin{equation}
F (\mathbf q)=\frac{e^{-i\mathbf q\cdot \mathbf
R}}{[1+(a^2q^2)/4]^2}[\cos (\mathbf q\cdot \mathbf L/2)I+i\sin
(\mathbf q\cdot \mathbf L/2)\sigma _z].
\end{equation}
The interaction can then be expressed in the form (\ref{int}), but
with different coupling constants,
\begin{equation}\label{hyg}
g_{\mathbf q}=i\Xi q \left( \frac \hbar   {2\rho qsV}\right)
^{1/2}\!\!\frac{e^{-i\mathbf q\cdot \mathbf R}}{[1+(a^2q^2)/4]^2}\sin
(\mathbf q\cdot \mathbf L/2).
\end{equation}

We note that the details of the electronic band structure, found important in
recent decoherence studies \cite{HU} in other contexts, would matter for
calculations that either consider decoherence due to processes mediated fully or
partially by interactions of the bound electron with the conduction electrons, or
if the impurity-bound electron were hybridized with the conduction electrons,
namely if its wave function were affected by the lattice potential outside the dots.
In our case, both effects are not considered. Specifically, we assume that the quantum
dots are strongly binding and that the electron wave functions are approximately Gaussian
or Hydrogen-like. While we do not expect the results to be modified significantly,
consideration of the band structure effects could be an interesting future project
especially for P-impurities in Si, where the bound-electron wave function actually
extends well beyond the lattice spacing.

\subsection{Error Estimates During Gate Functions}\label{Subsecti34}

In general, the ideal qubit evolution governed by the Hamiltonian term (\ref{He})
is time dependent. Decoherence estimates for some solid-state systems with
certain shapes of time dependence of the system Hamiltonian were reported recently
\cite{Brandes,Brandes2,Solenov}. However, such calculations are rather complicated.
Actually, there is no need to consider all possible time dependent controls of qubit
to evaluate its performance. All single-qubit rotations which are required for
quantum algorithms can be successfully implemented
by using two constant-Hamiltonian gates, e.g., amplitude rotation
and phase shift \cite{Kitaev3}.
To perform both of these gates one can keep the Hamiltonian term (\ref{He})
constant during the implementation of each gate,
adjusting the parameters $\varepsilon _{A}$ and $\varepsilon _{P}$
as appropriate for each gate and for the idling qubit in between gate functions.

In the following subsections we give specific examples: In Subsection \ref{NOT},
we will consider decoherence during the implementation of
the NOT gate (an amplitude gate). A $\pi$-phase shift gate is considered in
Subsection \ref{PI}. Then, in Subsection \ref{Subsecti37} we discuss the overall noise
level estimate for a qubit subject to gate control.

\subsection{Relaxation During the NOT Gate}\label{NOT}

The quantum NOT gate is a unitary operator which
transforms the states $|0\rangle$ and $|1\rangle$ into each other.
Any superposition of $|0\rangle$ and $|1\rangle$ transforms accordingly,
\begin{equation}
    {\rm NOT} \left(x |0\rangle + y |1\rangle\right) = y |0\rangle + x |1\rangle .
\end{equation}
The NOT gate can be implemented by properly choosing $\varepsilon_A$ and $\varepsilon_P$
in the Hamiltonian term (\ref{He}).
Specifically, with constant
\begin{equation}
 \varepsilon_A=\varepsilon
\end{equation}
and
\begin{equation}
 \varepsilon_P=0,
\end{equation}
the ``ideal'' NOT gate function is carried out, with these interaction parameters, over the time interval $T_g = \tau$,
\begin{equation}
 \tau=\frac{\pi\hbar}{\varepsilon}.
\end{equation}

The dominant source of quantum noise for double-dot qubit subject to the NOT-gate type coupling,
is relaxation involving energy exchange with the phonon bath (i.e., emission and absorption of phonons).
In this case it is more convenient to study the evolution of the density matrix
in the energy basis, $\left\{ \left|+\right\rangle,\left|-\right\rangle\right\}$,
where
\begin{equation}
\left|\pm\right\rangle=\left(\left|0\right\rangle \pm
\left|1\right\rangle\right)/\sqrt{2}.
\end{equation}
Then, assuming that the time interval of interest is $[0,\tau]$, the qubit density matrix can be expressed \cite{therm2} as follows,
\begin{eqnarray} \label{rho_rel}
\rho(t)=\left(
\begin{array}{cc}
  \rho_{++}^{th}+\left[\rho _{++}(0)-\rho_{++}^{th}\right]e^{ - \Gamma t} &
  \rho _{+-}(0)e^{ - (\Gamma/2-i\varepsilon/\hbar ) t} \\\\
  \rho _{-+}(0)e^{ - (\Gamma/2+i\varepsilon/\hbar) t} & \rho_{--}^{th}+\left[\rho _{--}(0)-\rho_{--}^{th}\right]e^{ - \Gamma t}  \\
\end{array}
\right)\! .
\end{eqnarray}
This is the standard Markovian approximation for the evolution of the density matrix. For large times,
this evolution would result in the thermal state, with the off-diagonal density matrix elements decaying to zero, while the diagonal ones approaching the thermal values proportional to the Boltzmann factors corresponding to
the energies $\pm \varepsilon /2$. However, we are only interested in such evolution for a short time interval, $\tau$, of a NOT gate.
The rate parameter $\Gamma$ is the sum \cite{therm2} of the phonon emission rate, $W^{e}$, and absorption rate, $W^{a}$,
\begin{equation}\label{Gamma}
    \Gamma=W^{e}+W^{a}.
\end{equation}

The probability for the absorption
of a phonon due to excitation from the ground state to the upper level is
\begin{equation}
  w^{\lambda}=\frac{2 \pi}{\hbar}|\langle f|H_{ep}|i\rangle|^2 \delta(\varepsilon
  -\hbar s q),
\end{equation}
where $|i\rangle$ is the initial  state with the extra phonon with
energy $\hbar s q$ and $|f\rangle$ is the final  state, $\mathbf
q$ is the wave vector, and $\lambda$ is the phonon polarization. Thus, we have to calculate
\begin{equation}\label{W}
W^a=\sum\limits_{\bf q, \lambda}
w^{\lambda}=\frac{V}{(2\pi)^3}\sum\limits_{\lambda}\int d^3 q\,
w^{\lambda}.
\end{equation}
For the interaction (\ref{int}) one can derive
\begin{equation}
w^{\lambda}=\frac{2 \pi}{\hbar}|g_{\mathbf q,\lambda}|^2 N^{th}
\delta(\varepsilon-\hbar sq),
\end{equation}
where
\begin{equation}
N^{th}=\frac{1}{\exp(\hbar sq /k_B T)-1}
\end{equation}
is the phonon occupation number at temperature $T$, and $k_B$ is the Boltzmann constant.

For the piezoacoustic interaction, the coupling
constant in (\ref{p6}) depends on the polarization.
For longitudinal phonons, the polarization vector has Cartesian components, expressed in terms of the  spherical-coordinate angles,
\begin{equation}  \label{cpl1}
\xi_1^{\parallel}=e_1=\sin\theta \cos\phi, \quad \xi_2^{\parallel}=e_2=\sin\theta \sin\phi,\quad
\xi_3^{\parallel}=e_3=\cos\theta,
\end{equation}
where $e_j=q_j/q$.
For transverse phonons, it is convenient to define the two polarization vectors $\xi_i^{\perp1}$ and $\xi_i^{\perp2}$ to have
\begin{equation}  \label{cpl2}
\xi_1^{\perp1}=\sin\phi,  \quad \xi_2^{\perp1}=-\cos\phi, \quad
\xi_3^{\perp1}=0,
\end{equation}
\begin{equation}\label{cpl4}
\xi_1^{\perp2}=-\cos \theta\cos\phi,  \quad \xi_2^{\perp2}=-\cos
\theta\sin\phi, \quad \xi_3^{\perp2}=\sin \theta.
\end{equation}

Then for longitudinal phonons, one obtains \cite{dd_ieee}
\begin{eqnarray}
w^{\parallel}&=&\frac{\pi}{\rho s V q
}M^2e^{-a^2q^2/4}\\
&&\times\, 9\sin^4\theta\cos^2\theta\sin^2\phi\cos^2\phi\sin^2
(qL\cos\theta/2).\nonumber
\end{eqnarray}
For transverse phonons, one gets
\begin{eqnarray}\nonumber
w^{\perp1}&=&\frac{\pi}{\rho s V q
}M^2e^{-a^2q^2/4}(-2\sin\theta\cos^2\theta\sin\phi\cos\phi\\
&&+\sin^3\theta\cos\phi\sin\phi)^2 \sin^2(qL\cos\theta/2),
\end{eqnarray}
\begin{eqnarray}\nonumber
w^{\perp2}&=&\frac{\pi}{\rho s V q
}M^2e^{-a^2q^2/4}(-2\sin\theta\cos\theta\cos^2\phi\\
&&+\sin\theta\cos\theta\sin^2\phi)^2 \sin^2 (qL\cos\theta/2).
\end{eqnarray}
By combining these contributions and substituting them in (\ref{W}), we get
the probability of absorption of a phonon for all polarizations,
\begin{eqnarray}\label{r1}
 W_{\rm piezo}^a&=&\displaystyle\frac{M^2}{20\pi\rho s^2\hbar L^5 k^4}\frac{\exp{
 \left(-\frac{a^2 k^2}{2}\right)}}{\exp\left(\frac{\hbar s k}{k_B T}\right)-1}
 \\\nonumber
 &&\times\left\{\left(kL\right)^5+ 5 kL\left[ 2
\left(kL\right)^2 - 21\right]
\cos\left(kL\right)\right.\\
&&+\left.\nonumber 15\left[7 -
3\left(kL\right)^2\right]\sin\left(kL\right)\right\} ,
\end{eqnarray}
where
\begin{equation}
 k=\frac{\varepsilon}{\hbar s}
\end{equation} is the wave-vector of the absorbed
phonon.

For the deformation interaction (\ref{hyg}), one can obtain the following result,
\begin{equation}
w=\frac {\pi \Xi^2}{\rho sV}\frac{q}{[1+(a^2q^2)/4]^4}\sin^2
(\mathbf q\cdot \mathbf L/2) \delta(\varepsilon - \hbar s q).
\end{equation}
The total probability for a phonon absorption is
\begin{equation}\label{IDA}
W_{\rm deform}^a= \frac{\Xi^2}{4 \pi \rho s^2\hbar}\frac{k^3}{\left(1+a^2
k^2 / 4\right)^4}\frac{1-\sin (kL)/(kL)}{\exp\left(\frac{\hbar s
k}{k_B T}\right)-1}.
\end{equation}

Finally, the expressions for the phonon emission rates, $W^e$, can be obtained by multiplying the above
expressions, (\ref{r1}) and (\ref{IDA}), by $(N_{th}+1)/N_{th}$.

\subsection{Dephasing During a Phase Gate}\label{PI}

The $\pi$ gate is a unitary operator which does not
change the absolute values of the probability amplitudes of a qubit
in the superposition of the $|0\rangle$ and $|1\rangle$ basis states.
It changes the relative phase between the probability amplitudes. Specifically,
any superposition of $|0\rangle$ and $|1\rangle$ transforms according to
\begin{equation}
    {\Pi} \left(x |0\rangle + y |1\rangle\right) = x |0\rangle - y |1\rangle .
\end{equation}
Over a time interval $\tau$, the $\pi$ gate can be carried out with constant interaction parameters,
\begin{equation}
 \varepsilon_A=0
\end{equation}
and
\begin{equation}
 \varepsilon_P = \varepsilon = \frac{\pi\hbar}{\tau}.
\end{equation}

In \cite{dd}, double-dot qubit dynamics during
implementation of phase gates was considered. The relaxation dynamics
is suppressed during the $\pi$ gate, because there is no tunneling between the dots.
Quantum noise then results due to pure dephasing, i.e., via the decay of the
off-diagonal qubit density matrix elements, while the diagonal density matrix elements remain constant.
In the regime of pure dephasing, the qubit density matrix can be represented as \cite{basis,Palma}
\begin{equation}\label{mm0}
\rho(t) =\left(%
\begin{array}{cc}
  \rho _{00} {(0)}& \rho _{01} {(0)}e^{ - B^2(t)+i\varepsilon t/\hbar} \\\\
  \rho _{10} {(0)}e^{ - B^2(t)-i\varepsilon t/\hbar} & \rho _{11} {(0)} \\
\end{array}%
\right),
\end{equation}
with the spectral function,
\begin{eqnarray}  \label{cd02}\nonumber
B^2(t)&=&\frac{8}{\hbar^2} {\sum\limits_{\mathbf q, \lambda} }
\frac{\left| g_{\mathbf q, \lambda}\right| ^2}{\omega _q^2}\sin ^2%
\frac{\omega _qt}2\coth \frac{\hbar \omega _q}{2 k_B T}\\
&=&\frac{V}{\hbar^2 \pi^3}\int d^3 q \sum\limits_{\lambda}
\frac{\left| g_{\mathbf q, \lambda} \right| ^2}{q^2 s^2}\sin
^2\frac{qst}2\coth \frac{\hbar q s}{2 k_B T}.
\end{eqnarray}

For the piezoelectric  interaction, the coupling constant $g_{\mathbf q, \lambda}$ was obtained in
(\ref{p6}), and expression for the spectral function takes the form
\begin{eqnarray}  \label{cp1}\nonumber
B^2_{\rm piezo}(t)&=&\displaystyle \frac{M^2}{2\pi ^3\hbar \rho s^3
}\!\int_0^{\infty } q^2dq\int_0^{\pi} \sin \theta d\theta
\int_0^{2 \pi} d\varphi
\\
&&\times\sum\limits_{\lambda}\frac {(\xi^{\lambda}
_1e_2e_3+\xi_2^{\lambda}e_1e_3+\xi_3^{\lambda}e_1e_2)^2}{
q^3}\exp(-a^2q^2/2){}\nonumber{}
\\
&&\times\sin^2 (q L \cos\theta)\sin ^2\displaystyle \frac{qst} 2\coth
\displaystyle\frac{\hbar q s}{2 k_B T},
\end{eqnarray}
c.f.~(\ref{cpl1})-(\ref{cpl4}).
For the deformation interaction, we have the coupling
constant (\ref{hyg}), and the expression for the spectral function is given by
\begin{eqnarray}\label{BIDA}
B^2_{\rm deform}(t)&=&\displaystyle \frac{\Xi^2}{\pi^2 \hbar \rho
s^3}\!\int_0^{\infty } q^2dq\int_0^{\pi} \sin \theta d\theta{}\nonumber{}
\\&&\times
\frac{\sin^2 (q L \cos\theta)}
{q(1+(a^2q^2)/4)^4}\sin^2\displaystyle \frac{qst} 2\coth
\displaystyle\frac{\hbar q s}{2 k_B T}.
\end{eqnarray}

\begin{table}[b]
\renewcommand{\arraystretch}{1.3}
\caption{Qubit parameters}
\centering
\begin{tabular}{|c||c|c|}
\hline
Parameter & GaAs double-dot qubit& Si double-impurity qubit\\
\hline
\hline
$\rho$,  kg/m$^3$ & $5.31\times10^3$ & $2.33\times10^3$\\
\hline
$s$, m/s & $5.14\times10^3$ & $9.0\times10^3$ \\
\hline
$\Xi$, eV & 3.3 & --- \\
\hline
$e_{14}$, C/m$^2$ & --- & 0.16 \\
\hline
$\kappa$ & --- & 12.8 \\
\hline
$M$,  eV/m & --- & $e e_{14}/(\varepsilon_0 \kappa)$\\
\hline
$L$, nm & 50 & 50\\
\hline
$a$, nm & 25 & 3\\
\hline
\end{tabular}
\end{table}

\begin{figure}[t]
\includegraphics[width=11 cm] 
{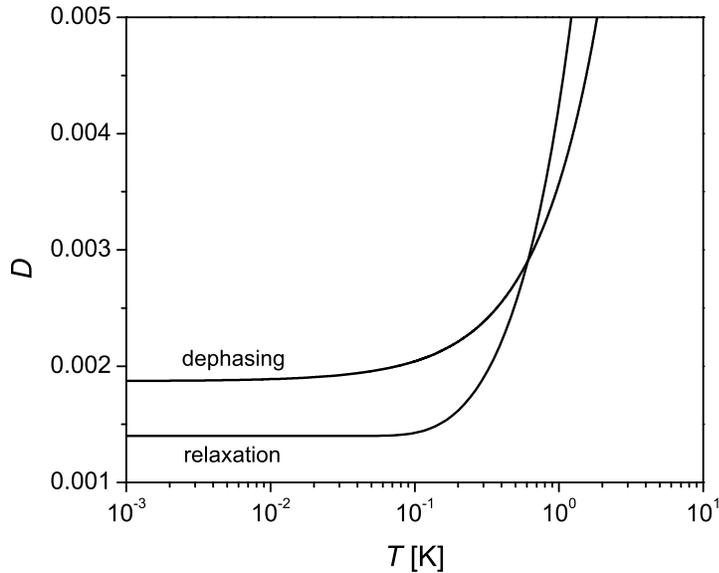} \caption{Estimates of the error measure per cycle, $D$, due to the
piezoelectric interaction in GaAs double-dot, shown as a
function of the temperature, $T$. The cycle time $\tau$ was $6\cdot 10^{-11}\,$s.}
\label{fig:2}
\end{figure}

\begin{figure}[t]
\includegraphics[width=11 cm] 
{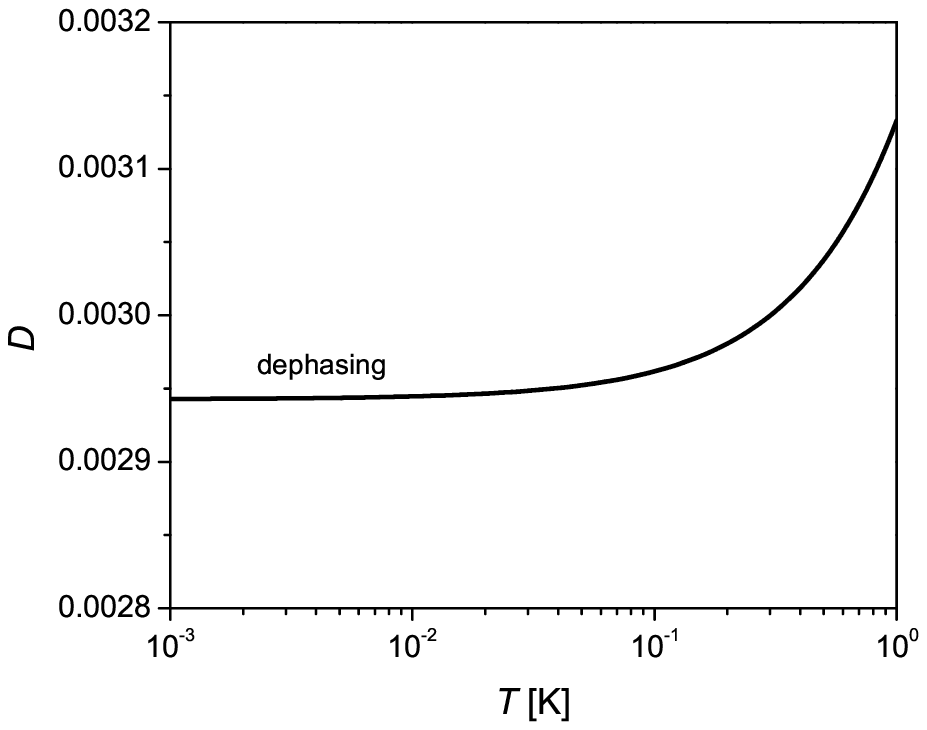} \caption{Estimate of error rate per cycle, $D$, due to
deformation phonon interaction for a double Phosphorus impurity in Si, shown as a
function of the temperature, $T$. The cycle time $\tau$ was $6\cdot
10^{-11}\,$s. The relaxation rate for this range of the parameter values is
negligibly small and respective values of $D$ are not shown.} \label{fig:3}
\end{figure}

\subsection{Qubit Error Estimates}\label{Subsecti37}

The qubit error measure, $D$, is obtained from the density matrix deviation from the ``ideal'' evolution
by using the operator norm approach \cite{norm} reviewed in Subsection \ref{Subsecti25}.
After lengthy
intermediate calculations one gets \cite{dd}
relatively simple expressions for the error during the NOT gate,
\begin{equation}\label{Da(t)}
 D_{\rm NOT}=\frac{1 - e^{ - \Gamma \tau}}{1+e^{ -\varepsilon /  k_B T  }},
\end{equation}
and the $\pi$ gate,
\begin{equation}\label{Dp(t)}
 D_{\pi}  = \frac{1}{2}\left[ 1 - e^{ - B^2 (\tau)}\right] .
\end{equation}

A realistic noise estimate could be taken as the worst case scenario, i.e.,
the maximum of these two expressions for error per gate cycle.
The expressions (\ref{Da(t)}) and (\ref{Dp(t)}) were used to calculate the
error rate for the double-dot qubit in GaAs and  double-impurity
qubit in Si. The parameters used were chosen to correspond to the experimentally realized
structures, \cite{Fujisawa,Hayashi,Dzurak1,Dzurak2}, and are summarized in Table 1.
The calculated error measures are presented in Figures 2 and 3.
The gate time $\tau$ selected for the reported calculations, $6\cdot
10^{-11}\,$s, is a representative value consistent with typical experimental conditions.
In fact, decreasing the gating time does not lead to smaller quantum noise in this case
because the energy gap of the driven qubit is $\sim 1/\tau $. If the gap
is made too large, other excitations will play a role in decoherence,
for instance, optical phonons. The time scale chosen here is within an optimal range, as discussed in \cite{dd}.

In summary, we derived expressions for the error measure for double-dot and
double-impurity qubits. The results, presented in Figures 2 and 3, suggest
that pure dephasing dominates at low temperatures. As the temperature increases
beyond about $1\,$K, the effect of relaxation becomes comparable and ultimately dominant.

The error measure values found, are 1.5 or more orders of magnitude larger than the
``traditional'' fault-tolerance thresholds for multiqubit quantum computation,
which range from $O(10^{-4})$ down to $O(10^{-6})$ \cite{Ben1,Ben2,Kitaev,Preskill,Knill98,Aliferis}.
However, recent developments have yielded less strict requirements for the error rate
\cite{Knill1,Knill2,Knill3}, optimistically, as large as
$O(10^{-2})$. Furthermore, there are several approaches to decrease decoherence effects
by pulsed control
\cite{Viola,Viola2,QCC,Byrd,Viola3,KK,Ozhigov,Viola4,KK2,Viola5},
some recently tested experimentally in multi-spin NMR \cite{Suter1,Suter2}.
Other ideas rely on the fact that instead of the bulk material, the
qubit could be manufactured in a one- or two-dimensional
nanostructure \cite{Dykman,Solenov2}, the latter already
available experimentally \cite{Eriksson}, which would affect the
phonon spectrum and lower decoherence effects.

\section{Additivity of Decoherence Measures}\label{Secti4}

In the study of decoherence of several-qubit systems,
additional physical effects should be taken into account.
Specifically, one has to consider the degree to which noisy environments of
different qubits are correlated \cite{Palma,JPCM}.
In addition to acting as a source of the quantum noise, the correlated bath
can induce an effective interaction, namely, create entanglement,
between the qubits immersed in it \cite{Solenov2,Dem,Solenov3,Solenov4}.
Furthermore, if all
constituent qubits are effectively immersed in the same bath, then
there are ways to reduce decoherence for this group of qubits
without error correction algorithms, by encoding the state of one logical qubit in a
decoherence-free subspace of the states of several physical
qubits \cite{Zanardi,Palma,DFS,Lidar,Wubs}.
In this section, we will consider several-qubit quantum registers
and, as the ``worst case scenario'' assume that the qubits experience
uncorrelated noise, i.e., each is coupled to a separate bath. Since
analytical calculations for several qubits are not feasible, we seek
``additivity'' properties that will allow us to estimate the error measure
for the register from the error measures of the constituent qubits.

It is important to emphasize that loss of quantum coherence results
in a loss of various two- and several-qubit entanglements in the system.
The highest order (multi-qubit) entanglements are ``encoded''
in the far off-diagonal elements of the multi-qubit register
density matrix, and therefore these quantum correlations will
decay at least as fast as the products of the decay factors for the qubits involved, as
exemplified by several explicit calculations
\cite{Eberly1,Storcz,Aravind,Eberly2}.
This observation leads to the conclusion that, for large times, the \emph{rates\/}
of decay of coherence of the qubits will be additive.

However, here we seek a different result: one valid not in the regime
of the asymptotic large-time decay of quantum coherence, but for relatively
short times, $\tau$, of quantum gate functions, when the noise level,
namely the value of the measure $D(\tau)$ for each qubit, is relatively small.
In this regime, we will establish \cite{additivity} in this section, that,
even for strongly entangled qubits$\,$---$\,$which is important for the utilization of the
power of quantum computation$\,$---$\,$the error measures $D$ of the individual qubits in a quantum register are additive.
Thus, the error measure for a register made of similar qubits, scales up linearly
with their number, consistent with other theoretical and experimental
observations \cite{Dalton,Suter1,Suter2}.

In Subsection \ref{D(t)}, we revisit the noise measure via the maximal
deviation norm and discuss some of its properties. In Subsection \ref{K(t)},
we introduce the diamond norm which is used as an auxiliary tool in the proof of additivity.
We then establish an approximate upper bound for $D(t)$ for a register of several weakly interacting
but possibly strongly entangled qubits, and cite work that further refines the additivity properties for typical qubit realizations.

\subsection{The Maximal Deviation Norm}\label{D(t)}

To characterize decoherence for an arbitrary initial state, pure
or mixed, we use the maximal norm, $D$, which was defined (\ref{normD})
in Subsection \ref{Subsecti25} as an operator norm maximized
over all the possible initial density matrices.
One can show that $0 \leq D(t) \leq 1$.  This measure of
decoherence will typically increase monotonically from zero at
$t=0$, saturating at large times at a value $D(\infty) \leq 1$.
The definition of the maximal decoherence measure $D(t)$ looks
rather complicated for a general multiqubit system. However, it can be
evaluated in closed form for
short times, appropriate for quantum computing, for a single-qubit
(two-state) system. We then establish an approximate additivity
that allows us to estimate $D(t)$ for several-qubit systems as
well.

In the superoperator notation the evolution of the reduced density
operator of the system (\ref{f1}) and the one for the ideal
density matrix (\ref{f2}) can be formally
expressed \cite{Kitaev,Kitaev2,Kitaev3} in the following way
\begin{equation}\label{T1}
\rho(t)=T(t)\rho(0),
\end{equation}
\begin{equation}\label{Ti}
    \rho^{(i)}(t)=T^{(i)}(t)\rho(0),
\end{equation}
where $T$, $T^{(i)}$ are linear superoperators. In this notation
the deviation can be expressed as
\begin{equation}\label{t1}
\sigma(t)=\left[T(t)-T^{(i)}(t)\right]\rho(0).
\end{equation}

The initial density matrix can always be written in the following
form,
\begin{equation}\label{mixture}
    \rho(0)=\sum_{j} p_j |\psi_j\rangle\langle\psi_j|,
\end{equation}
where $\sum_j p_j=1$ and $0 \leq p_j\leq1$. Here the set of the wavefunctions
$|\psi_j\rangle$ is not assumed to have any orthogonality properties.
Then, we get
\begin{equation}
\sigma\left(t, \rho(0)\right)=\sum_{j} p_j
\left[T(t)-T^{(i)}(t)\right]\left|\psi_j\right\rangle\left\langle\psi_j\right|.
\end{equation}
The deviation norm can thus be bounded,
\begin{equation}\label{proj}
\|\sigma(t, \rho(0))\|_{\lambda} \; \leq \;
  \left\|  \left[T(t)-T^{(i)}(t)\right]  |\phi\rangle\langle\phi|\right\|_{\lambda}.
\end{equation}
Here
$|\phi\rangle$ is defined according to
\begin{equation}\nonumber
\left\|  \left[T-T^{(i)}\right]  |\phi\rangle\langle\phi|\right\|_{\lambda}=
\max_j\left\|  \left[T-T^{(i)}\right]  |\psi_j\rangle\langle\psi_j|\right\|_{\lambda}.
\end{equation}
It transpires that for any initial density operator which is a
statistical mixture, one can always find a density operator which
is pure-state, $|\phi\rangle\langle\phi|$, such that $\|\sigma(t,
\rho(0))\|_{\lambda}\leq\|\sigma(t,
|\phi\rangle\langle\phi|)\|_{\lambda}$. Therefore, evaluation of
the supremum over the initial density operators in order to find
$D(t)$, see (\ref{normD}), can be done over only pure-state
density operators, $\rho (0)$.

Let us consider strategies of evaluating $D(t)$ for a single qubit. We can
parameterize $\rho(0)$ as
\begin{equation}\label{parametriazation1}
  \rho(0)=U \left(
\begin{array}{cc}
  P & 0 \\
  0 & 1-P \\
\end{array}
\right)U^{\dagger},
\end{equation}
where $0\leq P \leq 1$, and $U$ is an arbitrary $2 \times 2$
unitary matrix,
\begin{equation}
U=\left(
\begin{array}{cc}
  e^{i(\alpha+\gamma)}\cos\theta & e^{i(\alpha-\gamma)}\sin\theta
\\
  -e^{i(\gamma-\alpha)}\sin\theta & e^{-
i(\alpha+\gamma)}\cos\theta \\
\end{array}\right).
\end{equation}
Then, one should find a supremum of the norm of deviation
(\ref{n11}) over all the possible real parameters $P$, $\alpha$,
$\gamma$ and $\theta$. As shown above, it
suffices to consider the density operator in the form of a
projector and put $P=1$. Thus, one should search for the maximum
over the remaining three real parameters $\alpha$, $\gamma$ and
$\theta$.

Another parametrization of the pure-state density operators,
$\rho(0)=|\phi\rangle\langle\phi|$, is to express an
arbitrary wave function $|\phi\rangle=\sum_j (a_j+i b_j)|j\rangle$
in some convenient orthonormal basis $|j\rangle$, where $j=1,\ldots,N$.
For a two-level system,
\begin{equation}\label{parametrization2}
    \rho(0)=\left(%
\begin{array}{cc}
  a_1^2+b_1^2 &  (a_1+i b_1)(a_2-i b_2) \\
 (a_1-i b_1)(a_2+i b_2) &  a_2^2+b_2^2 \\
\end{array}%
\right),
\end{equation}
where the four real parameters $a_{1,2},b_{1,2}$ satisfy
$a_1^2+b_1^2+a_2^2+b_2^2=1$, so that the maximization is again
over three independent real numbers. The final expressions (\ref{Da(t)}) and
(\ref{Dp(t)}) for $D(t)$, for our selected single-qubit systems considered
in Section \ref{Secti3}, are actually quite compact and tractable.

In quantum computing, the error rates can be significantly reduced
by using several physical qubits to encode each logical
qubit \cite{Zanardi,DFS,Lidar}. Therefore, even before active
quantum error correction is
incorporated \cite{Ben1,Ben2,qec,Steane,Bennett,Calderbank,SteanePRA,Gottesman,Knill},
evaluation of decoherence of several qubits is an important, but
formidable task. Thus, our aim is to prove the approximate
additivity of $D_q(t)$, including the case of the initially strongly
\emph{entangled\/} qubits, labeled by $q$, whose dynamics
is governed by
\begin{equation}
  H=\sum_q H_q=\sum_q \left(H_{Sq}+H_{Bq}+H_{Iq}\right),
\end{equation}
where $H_{Sq}$ is the Hamiltonian of the $q$th qubit itself,
$H_{Bq}$ is the Hamiltonian of the environment of the $q$th qubit,
and $H_{Iq}$ is corresponding qubit-environment interaction. In the next subsection we consider a more complicated
(for actual evaluation) diamond norm
\cite{Kitaev,Kitaev2,Kitaev3}, $K(t)$, as an auxiliary quantity
used to establish the additivity of the more easily calculable
operator norm $D(t)$.

\subsection{Upper Bound for Measure of Decoherence}\label{K(t)}

The establishment of the upper-bound estimate for the maximal
deviation norm of a multiqubit system, involves several steps.
We derive a bound for this norm in terms of the recently
introduced (in the context of quantum computing) \cite{Kitaev,Kitaev2,Kitaev3}
diamond norm, $K(t)$.  Actually, for
single qubits, in several models the diamond norm can be
expressed via the corresponding maximal deviation norm. At
the same time, the diamond norm for the whole quantum system is
bounded by sum of the norms of the constituent qubits by using a
specific stability property of the diamond norm.  The use
of the diamond norm was proposed in \cite{Kitaev,Kitaev2,Kitaev3},
\begin{equation}\label{supernormK}
K(t) =\|T- T^{(i)}\|_{\diamond}=\sup_{\varrho} \|
\{[T-T^{(i)}]{\raise2pt\hbox{$\scriptscriptstyle{\otimes}$}} I\}
{\varrho} \|_{\rm Tr}.
\end{equation}
The superoperators $T$, $T^{(i)}$ characterize the actual and
ideal evolutions according to (\ref{T1}), (\ref{Ti}). Here $I$ is the
identity superoperator in a Hilbert space $G$ whose dimension is
the same as that of the corresponding space of the superoperators
$T$ and $T^{(i)}$, and $\varrho$ is an arbitrary density operator
in the product space of twice the number of qubits.

The diamond norm has an important stability property,
proved in \cite{Kitaev,Kitaev2,Kitaev3},
\begin{equation}\label{stability}
\|B_1 {\raise2pt\hbox{$\scriptscriptstyle{\otimes}$}}
B_2\|_{\diamond}=\|B_1\|_{\diamond} \|B_2\|_{\diamond}.
\end{equation}
Note that (\ref{stability}) is a property of the superoperators rather than
that of the operators.

Consider a composite system consisting of the two
subsystems $S_1$, $S_2$, with the noninteracting Hamiltonian
\begin{equation}
H_{S_1S_2}=H_{S_1}+H_{S_2}.
\end{equation}
The evolution superoperator of the system will be
\begin{equation}
 T_{S_1S_2}=T_{S_1}{\raise2pt\hbox{$\scriptscriptstyle{\otimes}$}}
T_{S_2},
\end{equation}
and the ideal one
\begin{equation}
T_{S_1S_2}^{(i)}=T_{S_1}^{(i)}{\raise2pt\hbox{$\scriptscriptstyle{\otimes}$}}
T_{S_2}^{(i)}.
\end{equation}
The diamond measure for the system can be expressed as
\begin{eqnarray}
&&K_{S_1S_2}^{\vphantom{(i)}}=\|T_{S_1S_2}^{\vphantom{(i)}} -
T_{S_1S_2}^{(i)}\|_{\diamond}=
\|(T_{S_1}^{\vphantom{(i)}}-T_{S_1}^{(i)}){\raise2pt\hbox{$\scriptscriptstyle{\otimes}$}}
T_{S_2}^{\vphantom{(i)}}+T_{S_1}^{(i)}{\raise2pt\hbox{$\scriptscriptstyle{\otimes}$}}
(T_{S_2}^{\vphantom{(i)}}-T_{S_2}^{(i)})\|_{\diamond}\nonumber\\
&&\leq\|(T_{S_1}^{\vphantom{(i)}}-T_{S_1}^{(i)}){\raise2pt\hbox{$\scriptscriptstyle{\otimes}$}}
T_{S_2}^{\vphantom{(i)}}\|_{\diamond}+\|T_{S_1}^{(i)}{\raise2pt\hbox{$\scriptscriptstyle{\otimes}$}}
(T_{S_2}^{\vphantom{(i)}}-T_{S_2}^{(i)})\|_{\diamond} . \label{justbelow}
\end{eqnarray}
By using the stability property (\ref{stability}), we get
\begin{eqnarray}
K_{S_1S_2}^{\vphantom{(i)}}\leq\|(T_{S_1}^{\vphantom{(i)}}-T_{S_1}^{(i)}){\raise2pt\hbox{$\scriptscriptstyle{\otimes}$}}
T_{S_2}^{\vphantom{(i)}}\|_{\diamond}+\|T_{S_1}^{(i)}{\raise2pt\hbox{$\scriptscriptstyle{\otimes}$}}
(T_{S_2}^{\vphantom{(i)}}-T_{S_2}^{(i)})\|_{\diamond}=\cr
\|T_{S_1}^{\vphantom{(i)}}-T_{S_1}^{(i)}\|_{\diamond}\|
T_{S_2}^{\vphantom{(i)}}\|_{\diamond}
+\|T_{S_1}^{(i)}\|_{\diamond}\|T_{S_2}^{\vphantom{(i)}}-
T_{S_2}^{(i)}\|_{\diamond}=\nonumber\\
\|T_{S_1}^{\vphantom{(i)}}-T_{S_1}^{(i)}\|_{\diamond}+\|T_{S_2}^{\vphantom{(i)}}-
T_{S_2}^{(i)}\|_{\diamond}=
K_{S_1}^{\vphantom{(i)}}+K_{S_2}^{\vphantom{(i)}}.
\end{eqnarray}

The inequality
\begin{equation}\label{Kbound}
 K\le \sum_q K_{q},
\end{equation}
for the diamond norm $K(t)$ has thus been obtained. Let us emphasize that
the subsystems can be initially entangled.
This property is particularly useful for quantum computing,
the power of which is based on qubit
entanglement. However, even in the simplest case of the
diamond norm of one qubit, the calculations are extremely
cumbersome. Therefore, the use of the measure $D(t)$ is preferable for actual
calculations.

For short times, of quantum gate functions, we can use (\ref{Kbound})
as an approximate inequality for order of magnitude estimates of decoherence measures,
even when the qubits are interacting. Indeed, for short times, the interaction effects
will not modify the quantities entering both sides significantly.
The key point is that while the interaction effects are small, this inequality can be
used for {\it strongly entangled\/} qubits.

The two deviation-operator norms considered are
related by the following inequality
\begin{equation}\label{a1}
\left\|   \sigma  \right\|_{{\lambda}}\leq\frac 1 2 \left\| \sigma
\right\|_{{\rm Tr}}\leq 1.
\end{equation}
Here the left-hand side follows from
\begin{equation}
\rm{Tr} \,\sigma =\sum_j\lambda_j =0.
\end{equation}
Therefore the $\ell$th eigenvalue of the deviation operator $\sigma$
 that has the maximum absolute value, $\lambda_\ell=\lambda_{\rm{max}}$, can be
expressed as \begin{equation}\lambda_{\ell}=-\sum_{j\neq
\ell}\lambda_j.\end{equation}
Thus, we have
\begin{equation}\label{a2}
 \left\|   \sigma  \right\|_{{\lambda}}=\frac 1 2\left(2 |\lambda_\ell|\right)
 \leq
 \frac 1 2\left(|\lambda_\ell|+\sum_{j\neq \ell}|\lambda_j|\right)=
\frac 1 2\left(\sum_j|\lambda_j|\right)=\frac 1 2\left\|\sigma
\right\|_{\rm Tr}.
\end{equation}
The right-hand side of (\ref{a1}) then also follows, because any density matrix has trace norm 1,
\begin{equation}\label{a3}
\|   \sigma  \|_{{\rm Tr}} = \|   \rho-\rho^{(i)} \|_{{\rm
Tr}}\leq \|   \rho \|_{{\rm Tr}}+ \|\rho^{(i)} \|_{{\rm Tr}}=2.
\end{equation}

From the relation (\ref{a3}) it follows that
\begin{equation}\label{prop}
 K(t)\le 2.
\end{equation}
 By taking the supremum of both sides of the relation (\ref{a2}) we get
\begin{equation}\label{prop1}
 D(t)=\sup_{\rho(0)}\left\|   \sigma  \right\|_{{\lambda}}\le
 \frac 12 \sup_{\rho(0)}\left\|   \sigma  \right\|_{\rm Tr}
 \le\frac 12 K(t),
\end{equation}
where the last step involves technical derivation details \cite{additivity} not reproduced here.
In fact, for a single qubit, calculations
for typical models \cite{additivity} give
\begin{equation}
 D_q(t)={\frac 1 2} K_q(t).
\end{equation}
Since $D$ is generally bounded by (or equal to) $K/2$,
it follows that the multiqubit norm $D$ is
approximately bounded from above by the sum of the single-qubit
norms even for the \emph{initially entangled\/} qubits,
\begin{equation}\label{DN1}
    D(t) \le \frac 12 K(t) \le \frac 12 \sum_q K_{q}(t)= \sum_q D_{q}(t),
\end{equation}
where $q$ labels the qubits.

For specific models of decoherence of the type encountered in Section \ref{Secti3},
as well as those formulated for general studies of short-time
decoherence \cite{norm}, a stronger property has been
demonstrated \cite{additivity}, namely that the noise measures are actually equal, for low levels of noise,
\begin{equation}\label{DN4-b}
    D(t)=\sum_q D_{q}(t)+o\left(\sum_q
D_{q}(t)\right).
\end{equation}

In summary, in this section we considered the maximal operator
norm suitable for evaluation of decoherence for a quantum register consisting of qubits immersed
in noisy environments. We established the additivity property of this measure of decoherence for
multi-qubit registers at short times, for which the level of quantum noise is low,
and the qubit-qubit interaction effects are small, but without any limitation on the
initial entanglement of the qubit register.

\section*{Acknowledgments}
We are grateful to A.~Fedorov, D.~Mozyrsky, D.~Solenov and D.~Tolkunov
for collaborations and instructive discussions. This research was supported by the
National Science Foundation, grant DMR-0121146.

\end{document}